# Probing defects in chemically synthesized ZnO nanostrucures by Positron Annihilation and Photoluminescence Spectroscopy

S K Chaudhuri[1,4], Manoranjan Ghosh[2,3,*], D Das[1] and A K Raychaudhuri[2]

[1]UGC-DAE Consortium for Research, III/LB-8, Salt Lake, Kolkata – 700 098, INDIA
[2]DST Unit for Nanoscience, S.N.Bose National Centre for Basic Sciences, Block-JD, Sector-3, Salt Lake, Kolkata-700 098, INDIA

**Abstract:** The present article describes the size induced changes in the structural arrangement of intrinsic defects present in chemically synthesized ZnO nanoparticles of various sizes. Routine X-ray diffraction (XRD) and Transmission Electron Microscopy (TEM) have been performed to determine the shapes and sizes of the nanocrystalline ZnO samples. Detailed studies using positron annihilation spectroscopy reveals the presence of zinc vacancy. Whereas analysis of photoluminescence results predict the signature of charged oxygen vacancies. The size induced changes in positron parameters as well as the photoluminescence properties, has shown contrasting or non-monotonous trends as size varies from 4 nm to 85 nm. Small spherical particles below a critical size (~ 23 nm) receive more positive surface charge due to the higher occupancy of the doubly charge oxygen vacancy as compared to the bigger nanostructures where singly charged oxygen vacancy predominates. This electronic alteration has been seen to trigger yet another interesting phenomenon, described as positron confinement inside nanoparticles. Finally, based on all the results, a model of the structural arrangement of the intrinsic defects in the present samples has been reconciled.

[3]Author's present address: Technical Physics Division, Bhabha Atomic Research Centre, Trombay, Mumbai-400 085, INDIA
[4] Author's present address: Department of Physics, University of Surrey, Guildford GU2 7XH, United Kingdom.

[*]Email: mghosh@barc.gov.in

## I. INTRODUCTION

Bulk ZnO crystal shows a sharp emission in the ultraviolet region due to excitonic recombination (exciton binding energy~60 meV) which makes ZnO a promising material for UV laser and light emitting diode (LED) at room temperature [1-4]. Interestingly, ZnO nanostructures also show a broad visible photoluminescence (PL) in the blue-green region (470 – 550 nm) which is linked to point defects in ZnO [5-7]. Different types of defects have been proposed to explain the two commonly observed defect emission bands in the visible range – a blue-green band ($\lambda_{bg}$ ~ 500-560 nm) and an orange-yellow band at ($\lambda_{oy}$ ~ 610 – 650 nm) [8,9]. The specific identification of the nature of defects for the emission in the visible region is difficult because of simultaneous existence of various types of defects in ZnO nanostructures [8,10]. Majority of the researchers believe that visible emission originates from oxygen vacancies located near to the surface region [10-12].

In one of our previous works [13], it has been shown that the emission energy as well as the intensity of the visible PL in ZnO nanostructures not only depend on the size, but are also related to their morphology. These nanostructures, synthesized by chemical route, were seen to undergo a shape transition at ~ 23 nm which changed the spectral components of the blue – green emission. In particular, spherical ZnO nanoparticles of size less than 23 nm showed broad visible photoluminescence around 550 nm (~2.2 eV) but the nanorods (of diameter > 23 nm and length ~ 100-150 nm) showed a peak around 500 nm (~2.5 eV). Both these emissions showed very systematic dependence on the surface to volume ratio which confirms the surface related origin of this emission [13,14]. It was also observed that change in emission energies and relative intensity of the spectral components due to shape transition is linked with the surface charge of the nanotructures which in turn depends on the ionic nature of the environment [15]. It is suggested that yellow (~2.2 eV) and green PL band (~2.5 eV) of ZnO nanostructures having different shapes and size originate from doubly ($V_0^{++}$) and singly ($V_0^{+}$) charged oxygen vacancy respectively [13,15]. Identification and attribution of defects present in ZnO nanostructures need additional support by complimentary technique like Positron Annihilation Lifetime Spectroscopy (PALS) frequently used for the investigation of defects present in semiconductor nanostructures [16].

In this paper, a sufficient number of nanocrystalline ZnO samples with different size and morphology have been investigated. The type of defects present in these nanostructures has been investigated by PL and PALS. It has been found that the visible photoluminescence characteristics as well as positron annihilation parameters are strongly correlated with the change in size of the nanoparticles. Possible explanation for the change in visible emission due to the change in different physical and environmental parameters has been discussed. With the help of PALS studies a structural defect arrangement in the nanostructures has been predicted which explains the mechanism of and some interesting features about the defect related visible emission. The positron annihilation lifetime data has revealed that positively charged oxygen vacancy at the surface creates a potential barrier which leads to the interesting phenomenon of confinement of positrons within the nanocrystals.

## II. SAMPLE PREPARATION AND EXPERIMENTAL TECHNIQUES

The controlled ZnO nanostructures (with wurtzite crystal structure) have been synthesized by acetate route described in detail elsewhere [17-19]. The smaller samples were prepared under ambient pressure as described in Ref 17. The larger samples (size > 23 nm) were prepared in an autoclave under 54 atm pressure at 230 $^0$C [13,18]. Precursor concentrations, reaction temperature and reaction time have been increased to synthesize larger particles. The details of representative nanostructures are given in Table 1. To synthesize the samples of different sizes, varied amount of $Zn(CH_3COO)_2.2H_2O$ has been reacted with NaOH of concentration three times that of the zinc acetate concentrations respectively for 0.5 hour to 6 hours duration. Two groups of samples have been investigated. Gr-I samples in Table 1 with spherical morphology (size < 23 nm) are synthesized under ambient pressure. Gr-II samples, having hexagonal and rod like shape (size > 23 nm) were synthesized in the autoclave.

For positron annihilation study, a 12 μCi $^{22}$Na activity deposited and sealed within two aluminium foils was used as the positron source. The PALS system used for this work is a standard fast-fast coincidence set-up using two identical 1-inch tapered off $BaF_2$ scintillator detectors fitted with XP2020Q photomultiplier tubes. The time resolution obtained using $^{60}$Co source with $^{22}$Na gates was 290 ps. A total of more than 1 million counts were recorded for each lifetime spectra. All the spectra were analyzed using PATFIT 88 [20]. Appropriate source correction has also been done. Each lifetime spectra were deconvoluted into three exponentially decaying lifetime components and a Gaussian resolution function to obtain the best fit. Several data were retaken and reanalyzed to monitor the stability of the spectrometer and reproducibility of the results.

Representative transmission electron microscope (TEM) images of the samples are shown in Fig. 1. It is seen that the samples with size below ~ 23 nm have a spherical shape. However, for size > 23 nm there is a shape transition and these samples show mostly hexagonal facets and some of them show rod like morphologies. Representative TEM images of spherical particles are shown in upper panel of Figs. 1(a)-1(c). The lower panel 1(d), 1(e), and 1(f) respectively shows rod like, hexagonal platelet and mixed morphologies (rod, cubes and square pillars) of ZnO nanostructures. Lattice image and SAED pattern of these nanostructures, not shown here, confirm their single crystalline nature. The crystal structures and size of all the samples have been analyzed by X-ray diffraction technique. X-ray diffraction data for the samples investigated are indexed in Fig. 2 which confirms wurtzite symmetry of the synthesized nanostructures. The size determined by TEM and XRD analysis (listed in table 1) agrees well within the range of ± 2 nm for particle size below 15 nm. The root mean square (rms) distribution in size (*d*) of the samples can be quantified as $\Delta d_{rms}/d \approx 10\%$ for d < 23 nm and $\Delta d_{rms}/d \approx 15\%$ for *d* > 23 nm respectively. The TEM data (Fig. 1 and Table 1) clearly establish that spherical ZnO nanoparticles prepared by chemical routes show a shape change to hexagonal shape with sharp edge (often rod like morphology) when the size exceeds the value of 23 nm.

# III. PROBING DEFECTS BY POSITRON ANNIHILATION SPECTROSCOPY

As mentioned in the experimental section, the positron lifetime spectra were deconvoluted into three components. In a nanocrystalline material with particle size smaller than the positron diffusion length (atmost 52 nm in ZnO [21]), the mechanism of positron annihilation may differ from that in single crystals [22]. Grain boundaries being rich in defects, act as strong trapping centers for positrons and a large fraction of positrons get trapped in the vicinity of grain boundaries rendering a few positrons to annihilate with free electrons inside a grain. So, in the case of nanocrystalline samples, the shortest lifetime $\tau_1$ and the corresponding percentage intensity $I_1$ is generally attributed to the annihilation events primarily at the structural defects in the grain interfaces. Hence, positron lifetime spectra of nanocrystalline materials generally have a little contribution from the electrons residing inside the grains and provide information regarding the annihilation at grain boundaries only. But in certain cases, also a fraction of positrons may cross more than one grain-surfaces and annihilate with free electrons inside a grain. In those cases, if the positron lifetime in the grain boundary defects is close to their free annihilation lifetime in defect free regions inside a grain, $\tau_1$ may represent a mixed lifetime comprising of both. The intermediate lifetime $\tau_2$ with corresponding intensity $I_2$ may be assigned to trapping of positrons in nano-voids at the intersection of three or more grain boundaries (e.g. triple junction) in nanocrystalline surface and the longer lifetime $\tau_3$ which has generally a smaller intensity $I_3$, is attributed to the pick-off annihilation of ortho-positronium formed in larger free volumes. The free volumes appear, most likely, because of the pores present in the pellets of the powdered sample and are not related to the sample property. Hence, the variation of $\tau_3$ and $I_3$ has not been discussed in the present case. Moreover, owing to its low intensity the effect of $\tau_3$ has not been included in the trapping model analysis as well.

A two-state trapping model, which assumes a presence of only one kind of vacancy type defect, can be applied to reveal some important facts about the annihilation mechanism of positrons [23]. In the case of two-state trapping model, the two lifetimes $\tau_1$ and $\tau_2$ are related by the following equations

$$\tau_1^{-1} = \tau_f^{-1} + K \qquad (1)$$

$$I_2 = 1 - I_1 = \frac{K}{K + \tau_f^{-1} - \tau_d^{-1}} \qquad (2)$$

where, $\tau_f$ is the lifetime corresponding to annihilation of positrons with free electrons in defect free regions and $\tau_d$ to that in the defect sites. K is the trapping rate of positrons in the defect site which is proportional to the concentration of positron traps. The value of $\tau_f$ is taken as 158 psec [24]. The trapping rate was calculated for different particle sizes using eqn. 2 with the experimentally obtained value of $\tau_2$ and $I_2$, and the value of $\tau_1$ according to the two-sate trapping model ($\tau_1^{TM}$) has been calculated using the K values in eqn. 1. Fig. 3 shows the comparison of the experimental lifetime $\tau_1$ with that calculated using the two-state trapping model. The mismatch between the two values readily suggests that a two-state trapping model is not valid which further indicates the presence of more than one type of defects. A consistently larger value of $\tau_1$ than that of $\tau_1^{TM}$ for all the samples suggests the presence of a defect having a lifetime close to $\tau_f$, which is difficult to resolve by a numerical fitting.

Chen et al [24] have reported that in ZnO, zinc vacancy ($V_{Zn}$) is the dominant defect which can be sensed by positrons, as they appear in negative and neutral charge states. The reported lifetime of positrons in Zn monovacancies is about 237 ps [25]. So, the shortest lifetime $\tau_1$ may be assumed to be a mixed state of positrons annihilating with free electrons in defect-free regions and zinc monovacancies. Now, if it is assumed that $n_f$ is the fraction of positrons annihilating in the defect free regions and $n_v$ is the fraction of positrons annihilating in the zinc vacancy sites with a lifetime $\tau_v$, then the experimentally obtained value of $\tau_1$ can be expressed as,

$$\tau_1 = n_f \tau_f + n_v \tau_v \qquad (3)$$

Fig. 4 and Fig. 5 give the variation of the lifetime components and the corresponding intensities respectively as a function of nanocrystal size. The value of $\tau_1$ was seen to increase with reduction in particle size from 85 nm to 23 nm. This increase in $\tau_1$ is quite obvious as the contribution of trapping at the zinc monovacancies in the interfaces increases due to the increase in the interface area with size reduction leading to an increase in $n_v$. Hence, from eqn. 3 it follows that $\tau_1$ will increase. A further reduction in the particle size should have increased the value of $\tau_1$. But enough interestingly, the value of $\tau_1$ was seen to decrease with decrease in particle size for samples having particle size smaller than 23 nm. In normal cases, a reduction in lifetime value suggests a decrease in the vacancy size. But as mentioned earlier, $\tau_1$ comprises of positron trapping in monovacancies and free annihilation. Since monovacancy is the lowest possible vacancy size, there is no scope of further size reduction. Also positron lifetime for free annihilation in a material does not change under normal circumstances. So, following eqn. 3, it can be said that the decrease in value of $\tau_1$ could only be due to an increase in the fraction $n_f$ or equivalently a reduction in the fraction $n_v$. It will be discussed in the next paragraph that the positron trapping rates at both the defect sites viz., monovacancies and the triple junction, decrease with size of the particle in the range of 5-23 nm. Hence, from the above arguments, it can be claimed that the possible reason behind the decrease in $\tau_1$ with decrease in particle size for samples having size below 23 nm, is a continuous increase in the contribution of free annihilation events which in turn is due to reduction of positron trapping at the defect sites.

The variation of defect concentration with particle sizes can be observed from the variation in trapping rates since the trapping rates are directly dependent on the concentration of defects. A three-state trapping model [26] can be applied to calculate the trapping rates at these defects. This model assumes that the shortest lifetime $\tau_1$ is an admixture of positron lifetime $\tau_{D1}$ in monovacancies and $\tau_f$, whereas $\tau_2$ (which is equal to $\tau_{D2}$ in the present trapping model) is the correctly resolved lifetime component. The trapping rates in both the defects were calculated using the following equations,

$$k_1 = \frac{\tau_1(\lambda_B - I_2\lambda_{D2}) - I_1}{\tau_{D1} - \tau_1} \qquad (4)$$

and

$$k_2 = \frac{I_2}{I_1}(\lambda_B - \lambda_{D2} + k_1) \qquad (5)$$

where the annihilation rates $\lambda_B = \tau_B^{-1}$ and $\lambda_{D2} = \tau_2^{-1}$. The lifetime $\tau_B$ is taken as 158 ps and $\tau_{D1}$ as 237 ps. Fig. 6 represents the variation of the trapping rates as a function of particle size. Both the trapping rates $k_1$ and $k_2$ were seen to decrease for particle sizes lower than 23 nm. All these observations support the fact that the contribution from the annihilation events at the grain boundaries is reducing as the particle sizes become smaller than 23 nm. From the above discussion it has been found that for the samples with average particle sizes lower than 23 nm, the contribution from the free annihilation events increases at the cost of annihilation events of the trapped positrons. As it will be seen in Section V, this anomalous behavior can be explained on the basis of size and morphology dependence of visible emission energy from ZnO nanostructures.

## IV. DEFECTS RESPONSIBLE FOR VISIBLE LUMINESCENCE IN ZnO NANOSTRUCTURES AND THEIR POSSIBLE LOCATION

It is a well accepted fact that presence of defects in nanostructured ZnO is manifested by their defect related visible emission in the blue-green region [5-8]. Photoluminescence results therefore can be employed to predict the defects responsible for controlling the optical and electronic properties of ZnO nanostructures. In this section the defects responsible for visible luminescence in ZnO nanostructures and their possible location has been investigated.

As can be seen from Fig. 7, size and morphology controlled ZnO nanostructures, in this study, show two line patterns in their room temperature luminescence spectra. The emission in the UV range (from 360-380 nm for sizes varying from 5 – 85 nm) is considered to be originated from the free exciton recombination [2] which is an intrinsic property of the wurtzite ZnO. Emission energy of this band edge emission shifts to the higher energies due to decrease in particle size due to quantum confinement. It can be seen from Fig. 8 that this energy shift obeys nearly 1/diameter dependence irrespective of the morphology and the method of sample preparation.

On the other hand, defect related emission in the visible region exhibits different behaviours for the two groups of samples of different morphology. Gr-I samples (size ranging from 5-15 nm) having spherical morphology emits in the yellow region (~550 nm) and the energy of this emission decreases from 529-564 nm as the particle size decreases (Fig. 7). As the flat faces appear for nanostructures greater than 23 nm (Gr-II samples), suddenly the position of the defect emission changes and occurs in the green region (500 nm). Surprisingly, the energy of this emission shifts to higher energies as particle size increases from (500 – 468 nm) which is exactly opposite in nature observed in case of Gr-I samples (Fig. 7). This anomalous size dependence in the defect related emission of the ZnO can be understood if the morphology of the nanostructures is taken into account. Non-spherical structures are formed due to widely different surface energies and the resulting anisotropy in the growth rate of ZnO crystal faces which eventually controls the optical and electronic properties of the nanostructures [13,15].

As shown in Fig. 7, the broad emission in the blue-green region is composed of two Gaussians marked as P1 and P2 [13, 27]. This two peak fitting procedure has a physical significance as far as the visible emission from ZnO is concerned. Role of the oxygen vacancy behind the broad visible emission was predicted in ref. 7 and 12. Origin of the two emission lines *P*1 and *P*2 has been described in ref. [6–12]. In ZnO nanocrystals the defect responsible for visible emission (wavelength range 470 - 650 nm) is the oxygen vacancy, denoted as $V_O$. The two lines originate depending on whether the emission is from doubly charged vacancy centre $V_o^{++}$ (*P*2) or singly charged vacancy centre $V_o^{+}$ (*P*1). The suggested emission process is explained by the following way. The $V_o^{++}$ centre, created by capture of a hole by the $V_o^{+}$ centre in a depletion region, leads to emission in the vicinity of 2.2 *eV* ($E_{P2}$), which is the *P*2 line [27]. The singly charged centre ($V_o^{+}$) in the absence of a depletion region becomes a neutral centre ($V_O^x$) by capture of an electron (*n*-type *ZnO*) from the conduction band which then recombines with a hole in the valence band giving rise to an emission (*P*1) at 2.5 *eV* ($E_{P1}$). Therefore, $V_o^{++}$ and $V_o^{+}$ were found responsible for emission near 2.2 eV (yellow band P2) and 2.5 eV (green band P1) respectively [13,27]. The shift in the visible emission is thus related to the change in the relative strengths of the two emission bands.

Another important issue is the location of the charged oxygen vacancies. Intensity ratio of NBE to visible emission ($I_{NBE}/I_{VIS}$) is an important parameter in quantifying the surface defect concentration responsible for the emission in the visible region. Increase in this ratio as the size increases predicts that visible emission centre is located predominantly at the surface (Fig. 9). A simple model [14] can be used to explain the intensity dependence on the size and thus obtain the thickness of the surface layer emitting visible light. The intensity ratio of the NBE to visible emission for cylindrical wires (Gr-II) of radius r and with a surface recombination layer of thickness t (Fig. 10), is given by [13,14]

$$\frac{I_{NBE}}{I_{VISIBLE}} = C\left(\frac{r^2}{2rt - t^2} - 1\right) \qquad (6)$$

By similar approach the same ratio can be found out for spherical particles (Gr-I) of radius r and a surface recombination layer thickness t (Fig. 10),

$$\frac{I_{NBE}}{I_{VISIBLE}} = C\left(\frac{r^3}{3rt(r-t) + t^3} - 1\right) \qquad (7)$$

Equation (6) and (7) have been employed to fit the experimental data for Gr-II and Gr-I samples respectively as shown in Fig. 9. The thickness of the surface recombination layers as obtained from the model is ~1.5 nm and ~3.6 nm for Gr-I (spherical) and Gr-II (rods and hexagonal platelets) particles respectively. Nanostructure surface, a crystal imperfection by definition is the likely sit of the defects and impurities that they receive during the method of preparation. Thus the size dependent optical properties are mainly dominated by the surface as the size of the nanostructures decreases.

# V. CORRELATION BETWEEN PHOTOLUMINESCENCE AND PALS RESULTS

It has already been seen that singly and doubly charged oxygen vacancies, located at the nanostructure surface, are responsible for 500 nm (P1) and 550 nm (P2) emission bands respectively. As described in Ref 13 the change in shape brings about a sharp change in the relative contribution of the two lines. P2 line is dominant in Gr-I samples (size ranging from 5-15 nm) having spherical morphology. As faceting starts appearing for nanostructures greater than 23 nm (Gr-II samples), intensity of P2 line gradually reduces and P1 emerges as the dominant emission band. The relative occupation of the two charged centres ($V_o^{++}$ and $V_o^{+}$) controls the surface charge and the depletion width of the nanostructures which in turn determines the occurrence of visible emission. Therefore, Gr-I spherical particles and Gr-II faceted samples are mostly occupied by the $V_o^{++}$ and $V_o^{+}$ respectively. So the spherical particles receive more positive charge than the rods and hexagonal platelets.

This fact has been confirmed by the Zeta potential measurement discussed in Ref 15. Both the samples receive positive charges due to the presence of charged oxygen vacancies but spherical particles of smaller size have higher positive charge (Zeta potential value is 18 mV) compared to the rods (Zeta potential value 8 mV). Results obtained from Energy Dispersive X-ray Spectroscopy (EDX) a surface sensitive tool for compositional analysis may be considered as relevant in this discussion. The weight ratio of O and Zn in stoichiometric ZnO should be 20% and 80 % respectively. But the weight ratio of oxygen determined from the average value of EDX data taken over different locations of the sample lies within the range 5.3 % to 10 % for the samples in our study. It proves that surfaces of the synthesized nanostructures in our study are oxygen deficient.

In the above discussion it has been seen that the spectral content or the relative weight of two lines (P1 and P2) of the blue-green band is determined by the relative occupancy of these charged oxygen vacancies. But the neutral oxygen vacancies ($V_O$) are invisible to positrons in the sense that positrons are not trapped in these defects since positron binding energy to neutral oxygen vacancy $V_O$ is only 0.04 eV [24]. Further addition of positive charge in the form of positively charged oxygen vacancies rules out the possibilities of positron trapping in $V_o^{++}$ and $V_o^{+}$. So, the zinc vacancies present in the sample are responsible for the trapping of positrons. Occurrence of zinc and oxygen vacancy in same place is mutually exclusive but they may co-exist in different physical segment of the sample although the samples are oxygen deficient as a whole. Now the positive charge at the sample surface created by $V_o^{++}$ and $V_o^{+}$ forms a potential barrier for the positrons. The barrier is higher in the case of small spherical particles (due to the presence of $V_o^{++}$) than the rods and hexagonal platelets (due to less positive $V_o^{+}$) of bigger size. As shown in Fig. 10, a simple un-scaled model is proposed where difference in barrier height for positrons in case of two different morphologies has been demonstrated. The positrons, which have initially got higher energies, manage to enter a particle but cannot escape after losing its energy through various inelastic collision processes. However a fraction of positrons still gets trapped at the zinc vacancies as can be seen from the difference in the $\tau_1$ and $\tau_1^{TM}$ (Fig. 3). As a result of increased number of positrons getting trapped inside the particle, the chances of their annihilation with free electrons in the defect free regions increases thereby increasing the fraction $n_f$ for small spherical particles. So, $\tau_1$ for these samples decreases. Therefore for samples with particle sizes between 85 nm to 23 nm, the zinc vacancies at the particle boundaries participate substantially to trap positrons because the trapping rate increases as the particle size decreases. But for the samples with average particle sizes below 23 nm the positive surface charge suddenly increases. This increase in the barrier height confines positrons within the bulk core of a nanocrystal which contains fewer defects (zinc vacancies) than the surface. So the trapping rate in zinc vacancies decreases for small spherical particles of size < 23 nm.

## VI. CONCLUSIONS

In this article simultaneous occurrence of various types of defects in ZnO nanostructures has been observed upon investigation using techniques like PALS and PL. The chemically prepared nanocrystals have been categorized into two groups. Gr-I samples of size 5 -15 nm showed spherical morphology whereas the Gr-II samples of size 23 – 100 nm were mostly rods and hexagonal platelet. It has been found that samples with spherical (Gr-I) morphologies show visible luminescence at 2.2 eV due to presence of doubly charged ($V_o^{++}$) and those with faceted (Gr-II) morphologies show a visible luminescence at 2.2 eV due to presence of singly charged oxygen vacancies ($V_o^+$) as well. On the other hand, the positron annihilation studies have suggested the presence of zinc vacancies in all the samples. Detailed analyses and reconciliation of all the experimental results has helped us to model the structural arrangement of defects in these nanocrystals. This model suggests that the Gr-I samples have excess of positive charge on their surface due to the presence of both $V_o^+$ and $V_o^{++}$. Potential barrier created by these positive charges is high enough to confine positrons which can enter such a nanocrystal once but eventually lose some of their energy by inelastic collisions. For the Gr-II samples, change in positron annihilation parameters with change in nanocrystalline size is quite normal and suggests presence of zinc vacancies apart from $V_o^+$ seen from PL studies.

TABLE 1: Size and morphology of ZnO nanostructures

| Identi-fication | | Morphology | Size by TEM (nm) | Size by XRD (nm) | $E_{NBE}$ (eV) |
|---|---|---|---|---|---|
| I | (i) | Spherical | 5 | 5 | 3.45 |
| | (ii) | Spherical | 10 | 10 | 3.35 |
| | (iii) | Spherical | 15 | 15 | 3.34 |
| II | (iv) | Rod diameter | 24 | 23 | 3.33 |
| | (v) | Hexagonal platelet | 45 | 44 | 3.28 |
| | (vi) | Rods and square pillars | 100-500 | 85 | 3.27 |

**FIGURE CAPTIONS**

FIG. 1. (a), (b) and (c) show TEM images of Gr-I spherical nanostructures of average size 5 nm, 10 nm and 15 nm resspectively. Lower panel shows the Gr –II samples of rod like [(d)], hexagonal platelets [(e)] and complex morphology with rods and square pillars [(f)].

FIG. 2. X-ray diffraction pattern for all the samples investigated show wurtzite symmetry (peaks are indexed). Gradual increase in the average size of the nanostructures can be seen from decreasing peak width.

FIG. 3. (Color online) Comparison of experimental value of $\tau_1$ and calculated value of $\tau_1$ from two-state trapping model, denoted by $\tau_1^{TM}$.

FIG. 4. Variation of measured lifetime values $\tau_1$ and $\tau_2$ as a function of nanostructure size.

FIG. 5. Variation of intensities $I_1$ and $I_2$ corresponding to positron lifetimes $\tau_1$ and $\tau_2$ respectively, as a function of nanostructure size.

FIG. 6. (Color online) Positron trapping rates $k_1$ and $k_2$ calculated by three state trapping model are plotted as a function of nanostructure size.

FIG. 7. (Color online) The emission bands of the ZnO nanostructures of different size and shape. The excitation was at 325 nm. The sharp dependence of the emission band at 490-565 nm on the size can be seen. A representative two Gaussian fitting has been shown for the 15 nm spherical nanoparticle.

FIG. 8. The Near Band Edge (NBE) emission energy has been plotted as a function of diameter of the nanostructures. An approximate 1/diameter dependence can be observed.

FIG. 9. Intensity ratio of NBE to visible emission for Gr-I and Gr-II samples. The data are fitted by a surface recombination model with surface layer thicknesses of 1.5 and 3.6 nm for Gr-I and Gr-II samples, respectively.

FIG. 10. (Color online) There exists a surface layer of thickness t where defects such as singly and doubly charged oxygen vacancies are mostly located. The horizontal arrows indicate the energy barrier seen by the positrons due to the presence of positively charged oxygen vacancies at the surface. Spherical particles (Gr-I) of size < 23 nm having higher positive surface charge offer higher barrier for positrons confined in the bulk core region than rods and hexagonal platelets (Gr-II) of size > 23 nm.

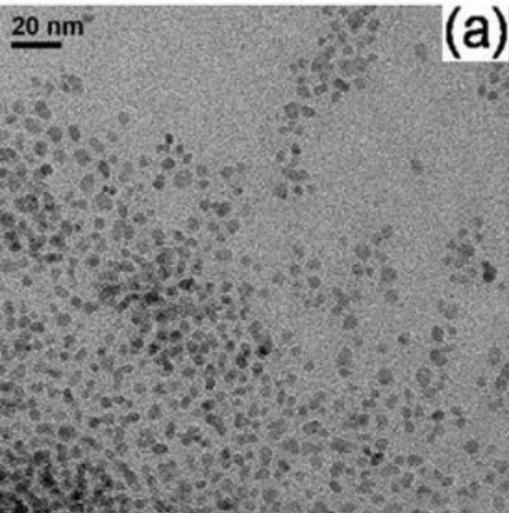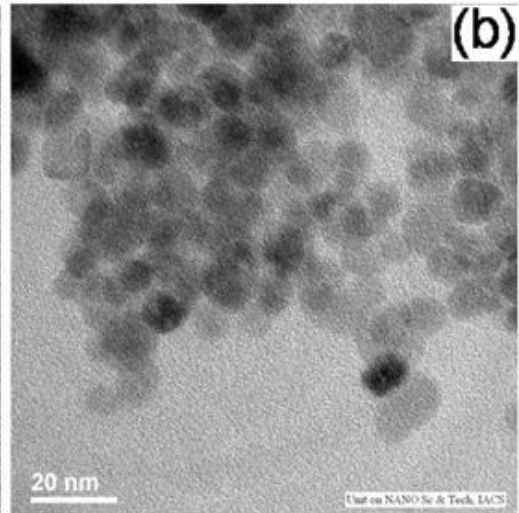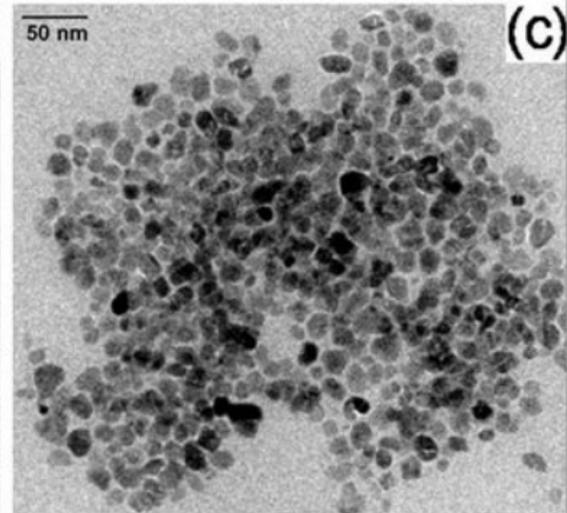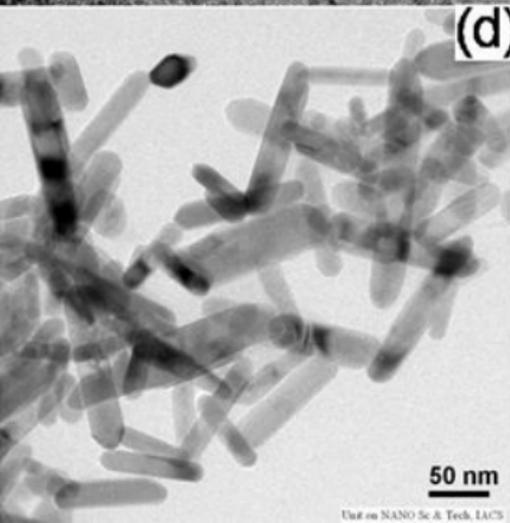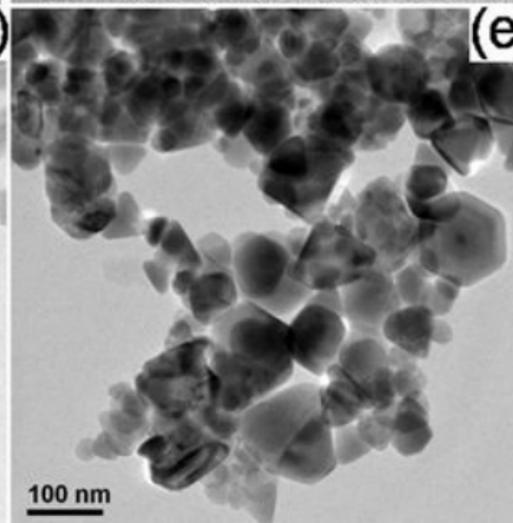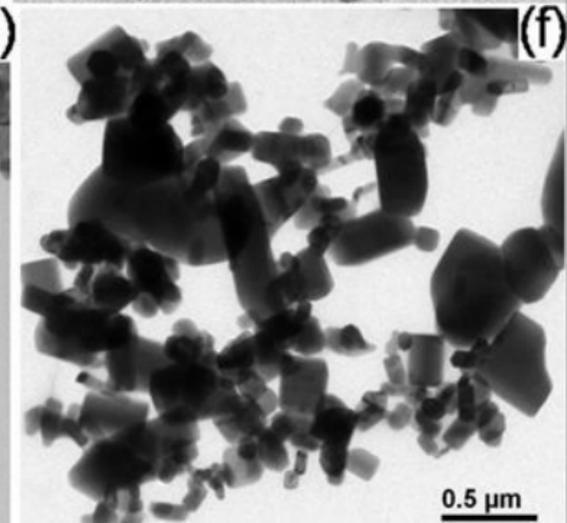

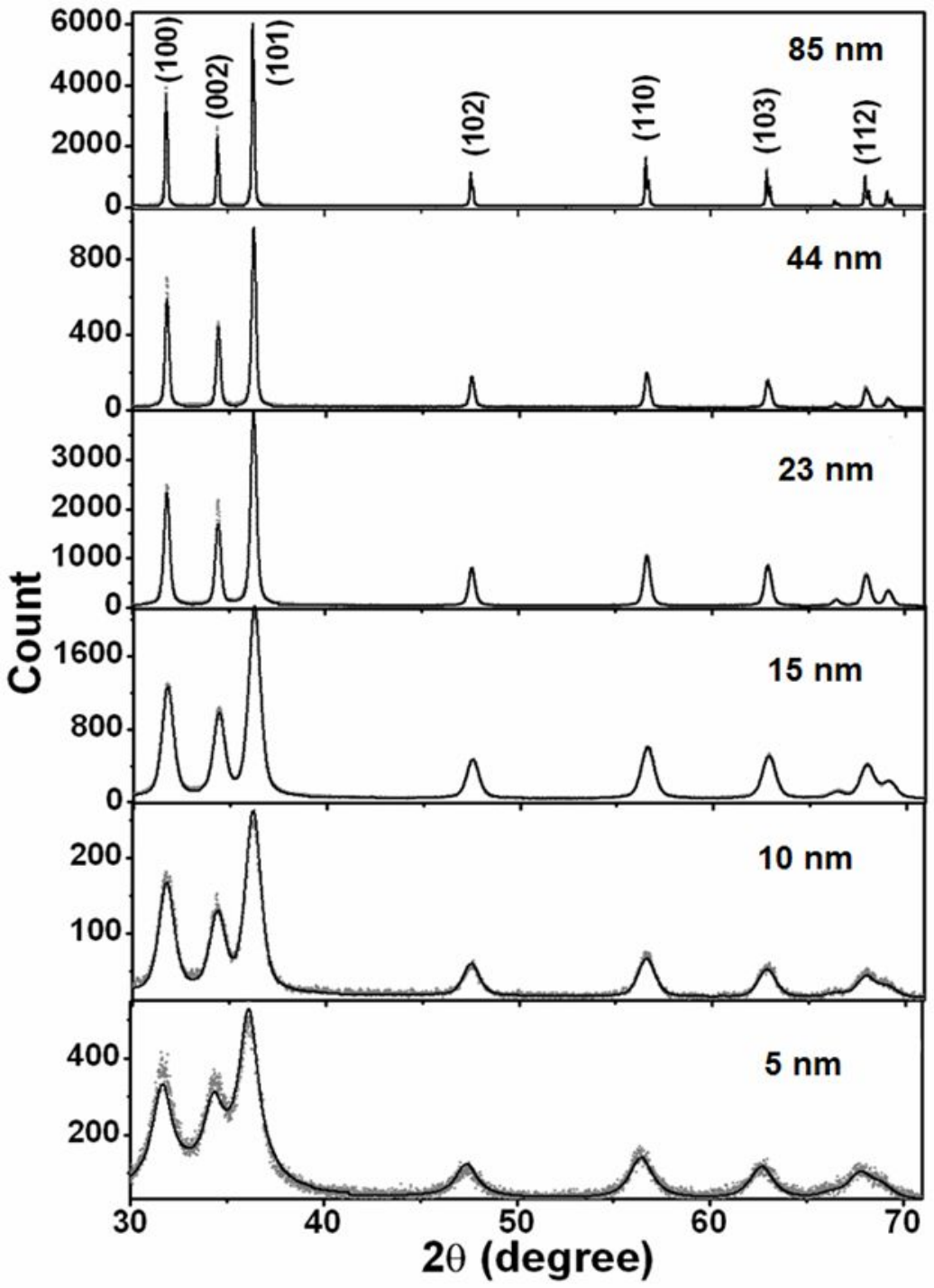

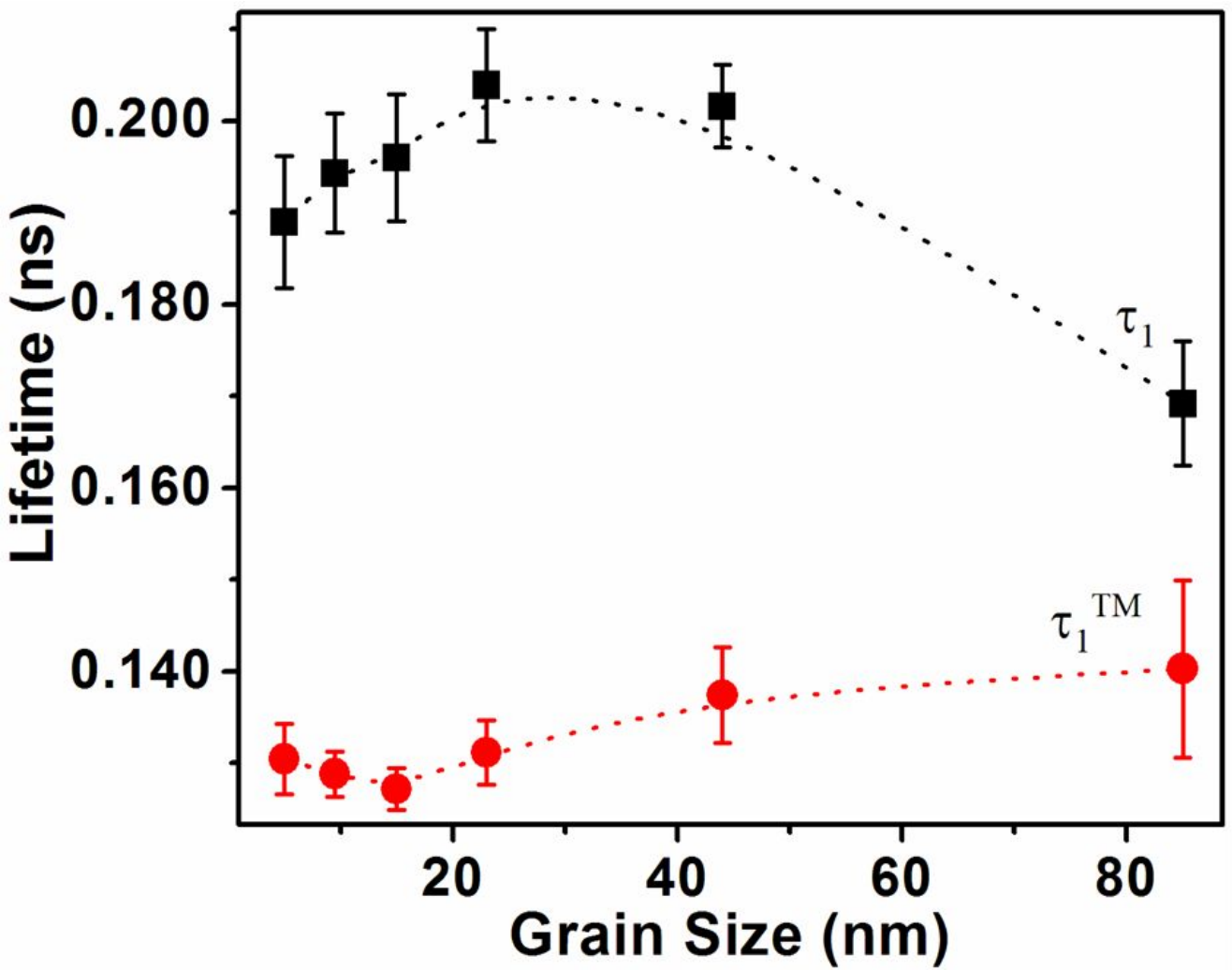

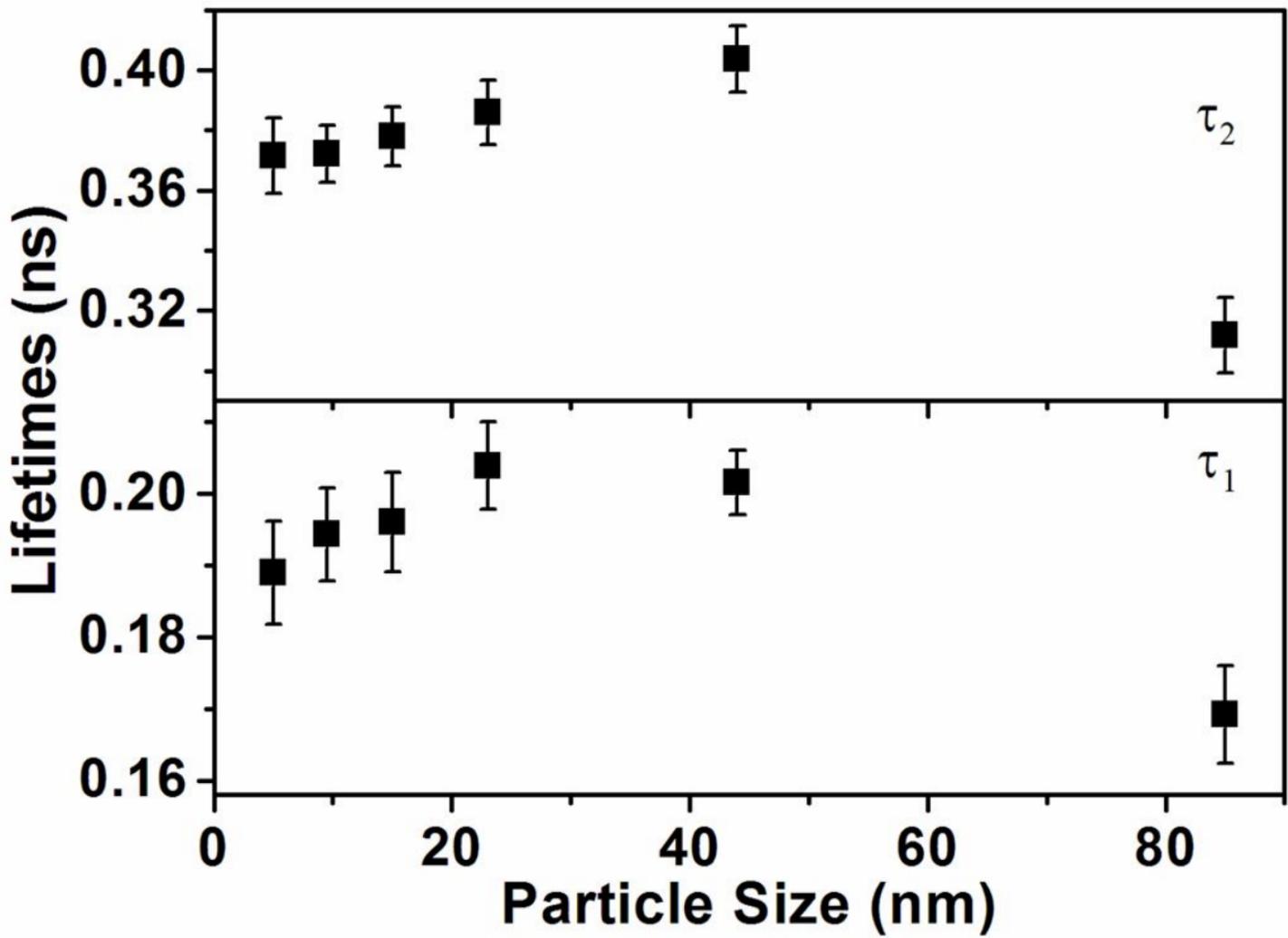

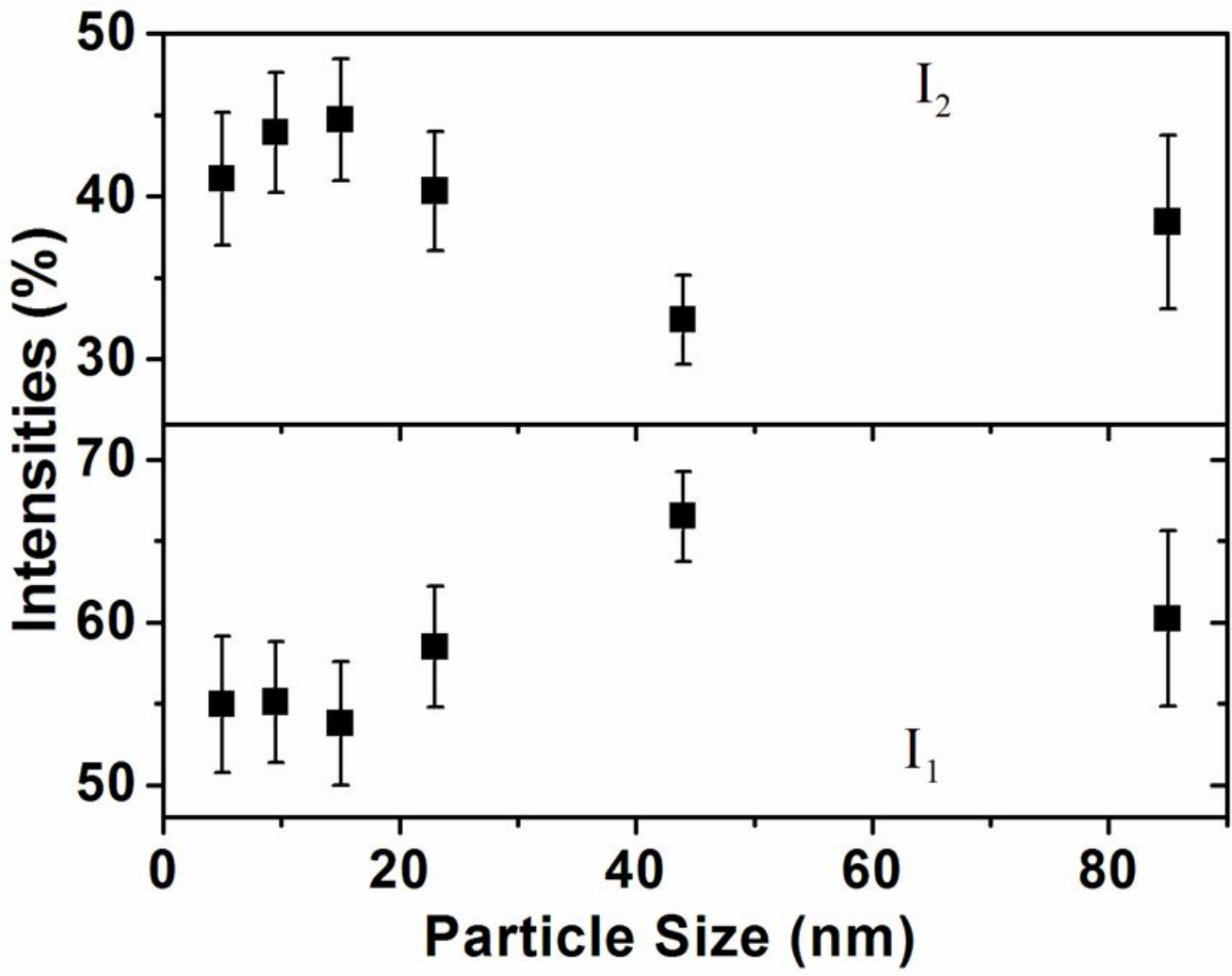

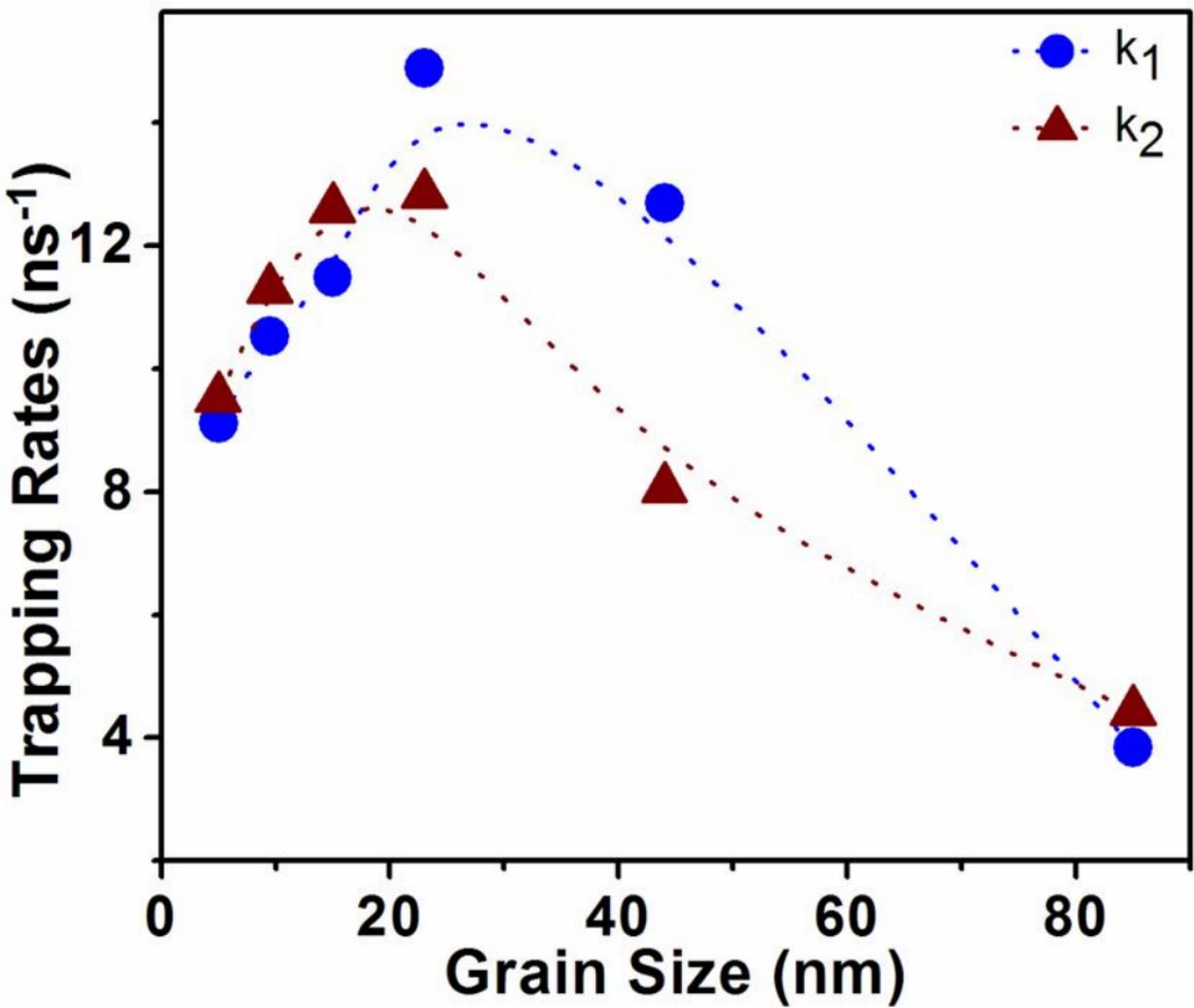

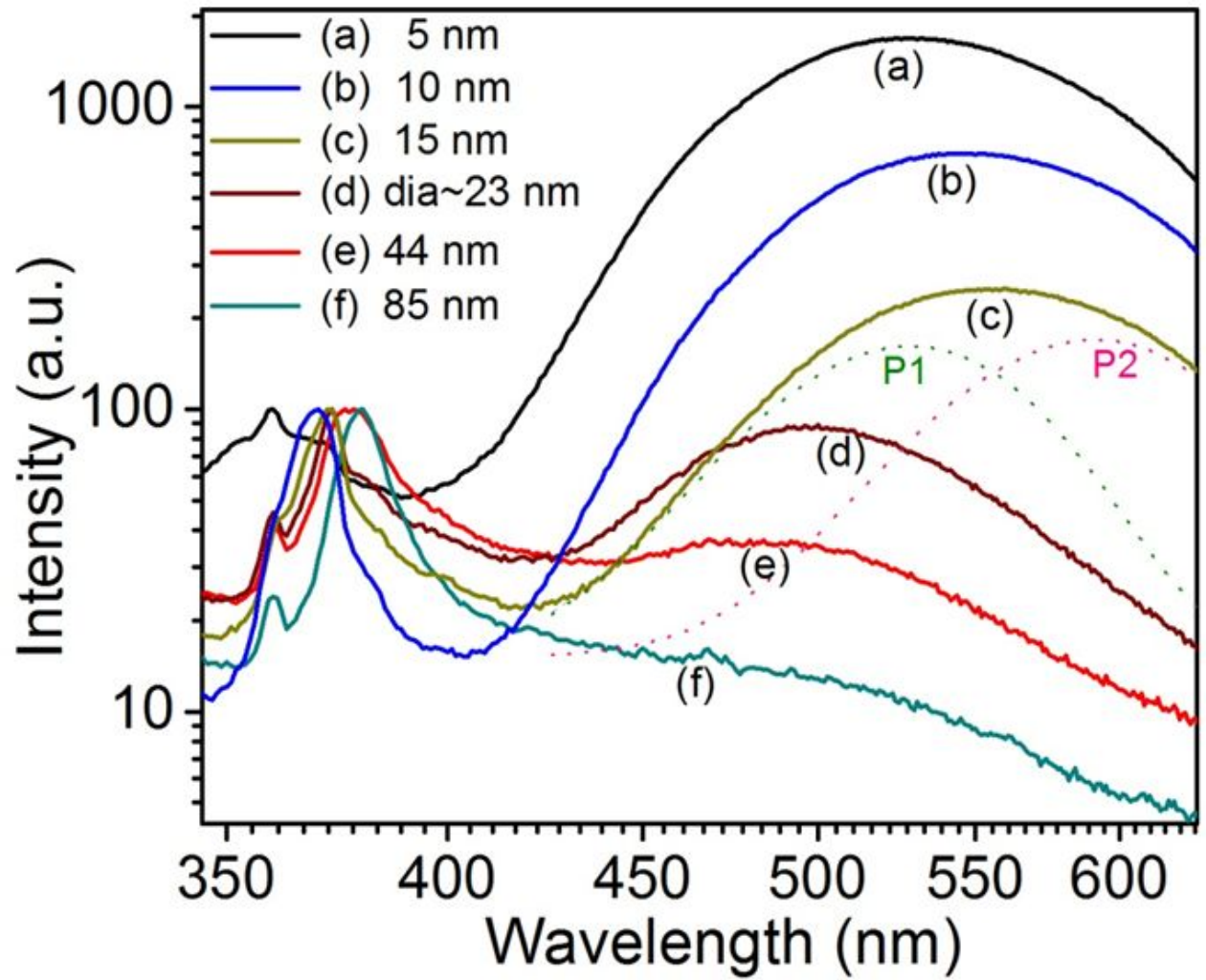

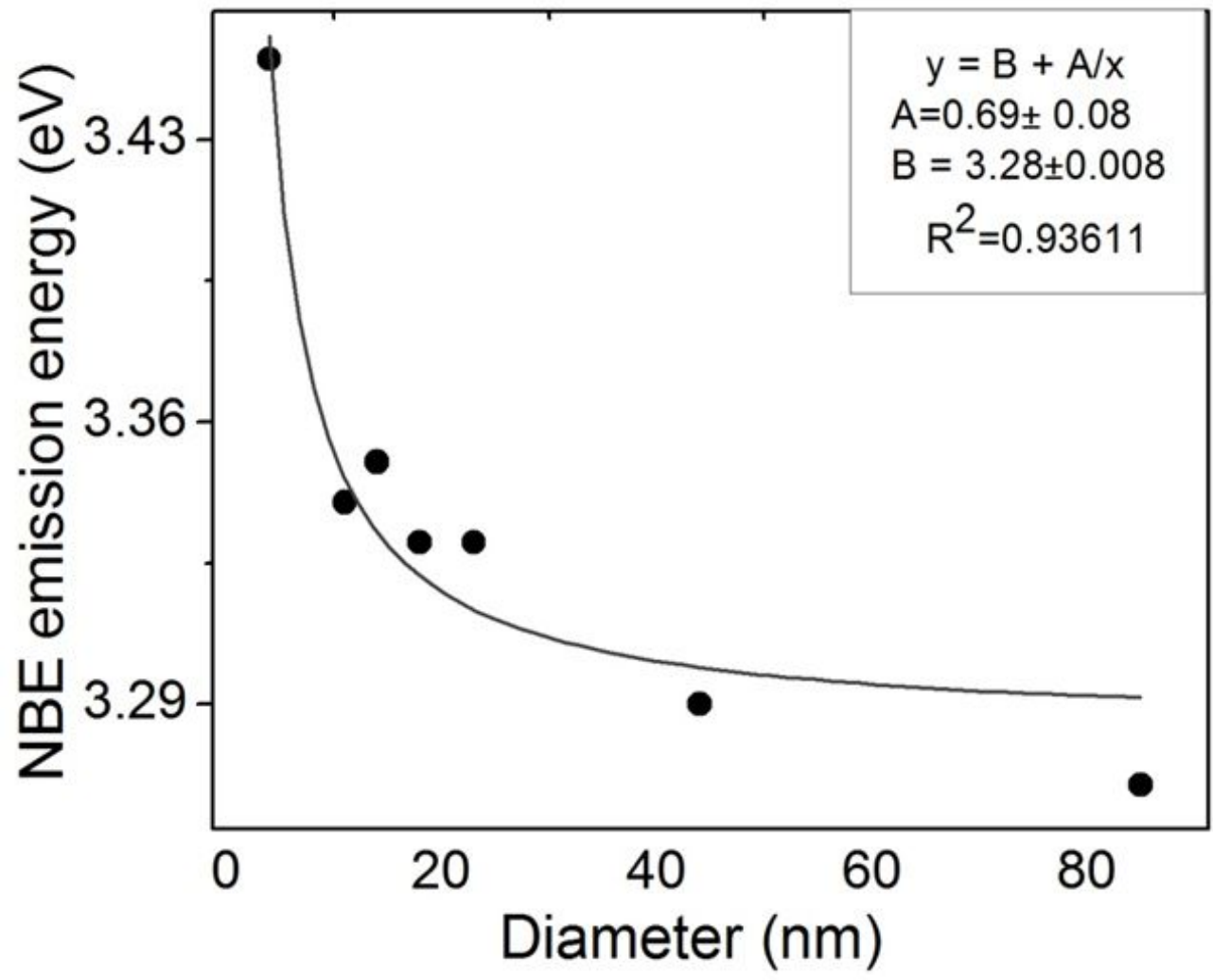

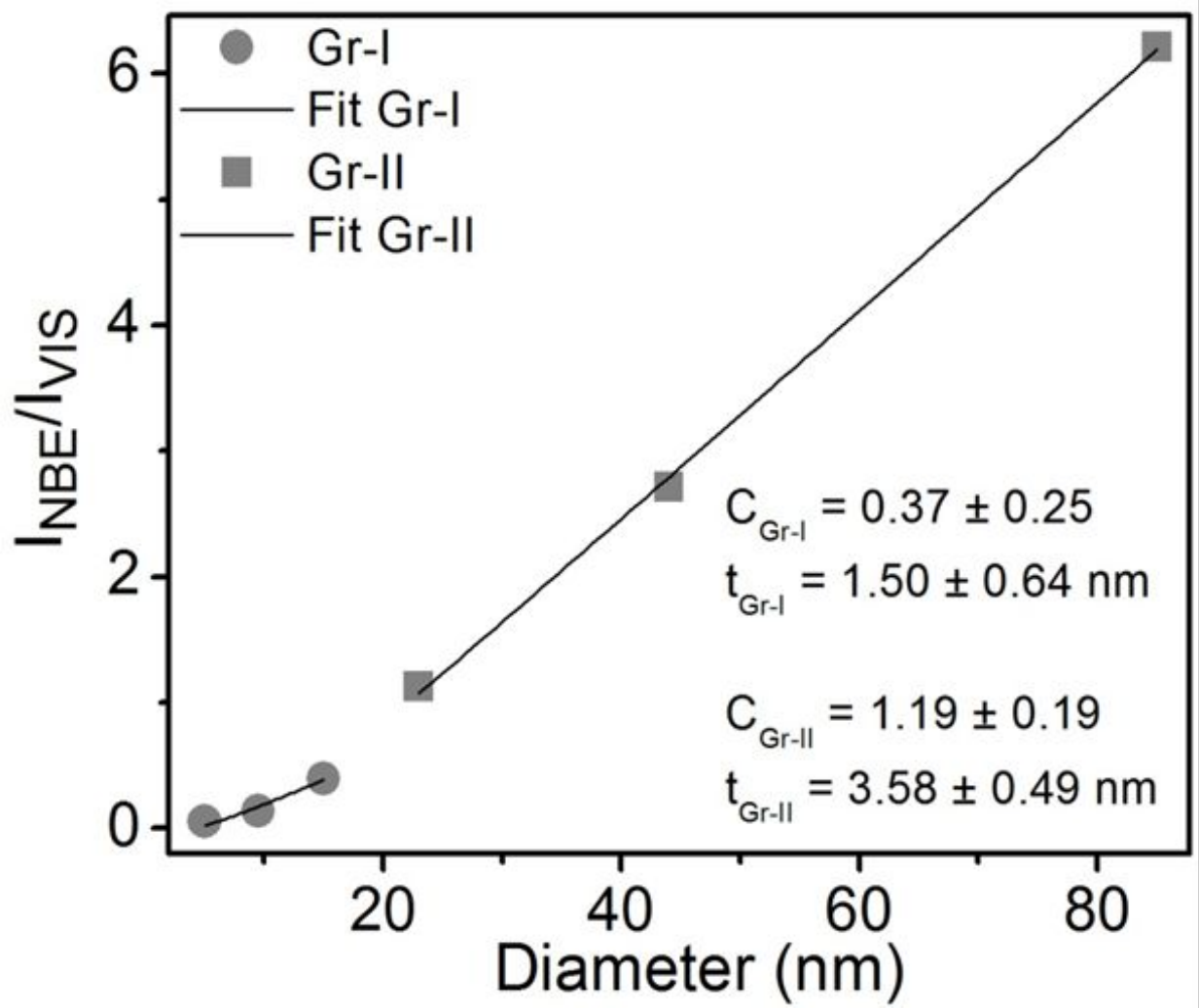

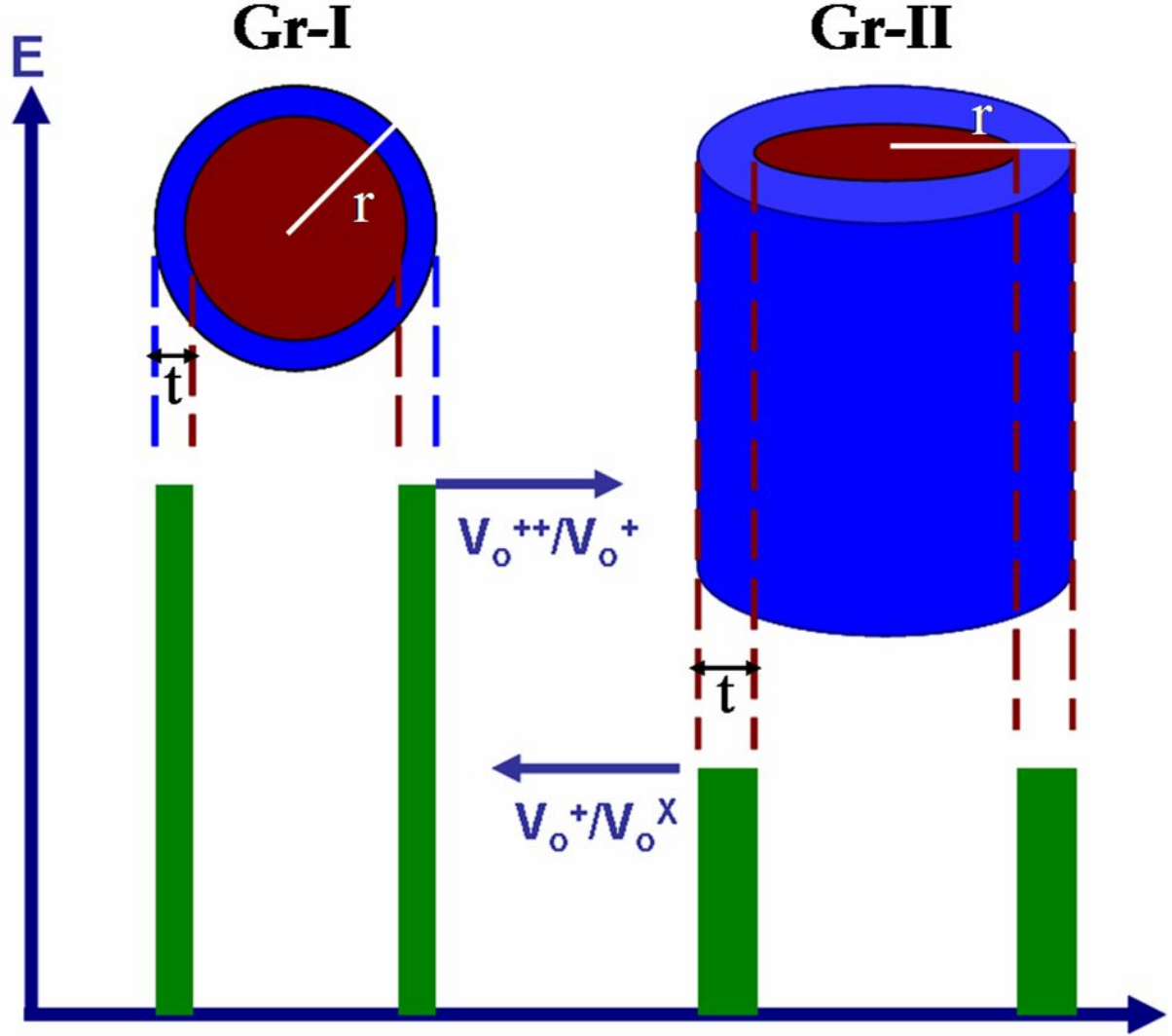